\documentclass[lettersize,journal]{IEEEtran}
\usepackage{amsmath,amsfonts}
\usepackage{algorithmic}
\usepackage{algorithm}
\usepackage{array}
\usepackage[caption=false,font=normalsize,labelfont=sf,textfont=sf]{subfig}
\usepackage{textcomp}
\usepackage{stfloats}
\usepackage{url}
\usepackage{verbatim}
\usepackage{graphicx}
\usepackage{cite}
\usepackage{microtype}
\usepackage{graphicx}
\usepackage{booktabs} 
\usepackage{multirow}
\usepackage{mathtools}
\usepackage{tipa}
\usepackage[table,xcdraw]{xcolor}
\usepackage{tabularx}
\usepackage[most]{tcolorbox}
\usepackage{array}
\usepackage{float}
\usepackage{enumitem}
\usepackage{pdfpages} 

\usepackage[hidelinks]{hyperref}
\usepackage{amsmath}
\usepackage{amssymb}
\usepackage{mathtools}
\usepackage{amsthm}

\begin{document}

\title{Commanding at the Speed of Sound: Audio-Centric Referring Expression Segmentation for Aerial Scenes}

\author{Ziyun Chen, Yijun Wang, Chuanyi Zhang, \IEEEmembership{Member,~IEEE}, Liang Yao,  Yuexuan An, \IEEEmembership{Member,~IEEE}, \\ Xiang Gu, Pai Peng, Wei Zhou, \IEEEmembership{Senior Member,~IEEE}, Fan Liu, \IEEEmembership{Senior Member,~IEEE}
\thanks{This work was supported in part by the National Natural Science Foundation of China under Grant 62372155, in part by Basic Research Program of Jiangsu under Grant BK20250188, in part by the China Postdoctoral Science Foundation under Grant Number 2025M771578, in part by the Research Funds of Jiangsu Hydraulic Research Institute under Grant 2025z065.
\textit{(Ziyun Chen and Yijun Wang contributed equally to this work.) (Corresponding author: Fan Liu (email: fanliu@hhu.edu.cn))}}
\thanks{
Ziyun Chen, Yijun Wang, Liang Yao, Yuexuan An and Xiang Gu are with the College of Computer Science and Software Engineering, Hohai University, Nanjing, 210098, China.

Chuanyi Zhang is with the College of Artificial Intelligence and Automation, Hohai University, Changzhou, 213200, China.

Pai Peng is with COWARobot Co. Ltd., Shanghai, 201805, China.

Wei Zhou is with the School of Computer Science and Informatics, Cardiff University, Cardiff CF24 4AG, United Kingdom.

Fan Liu is with the Key Laboratory of Water Big Data Technology of Ministry of Water Resources, College of Computer Science and Software Engineering, Hohai University, Nanjing, 210098, China.
}}

\markboth{Journal of \LaTeX\ Class Files,~Vol.~14, No.~8, August~2021}%
{Shell \MakeLowercase{\textit{et al.}}: A Sample Article Using IEEEtran.cls for IEEE Journals}


\maketitle

\begin{abstract}
Referring expression segmentation (RES) aims to localize targets at the pixel level guided by human instructions. However, current models predominantly follow a text-centric paradigm that treats discrete text as the primary interaction medium for these human instructions. This assumption is especially limiting in unmanned aerial vehicle (UAV) applications, where rapid response, hands-free operation, and natural communication are often critical. In this paper, we reconstruct UAV referring expression segmentation through an audio-centric perspective. Grounded in information theory, we demonstrate that raw audio provides a higher upper bound for extracting rich semantic features than transcribed text. To achieve this, we construct an audio-driven UAV referring expression segmentation benchmark named Aerial-AudioRES by augmenting VisDrone with synthesized speech, forming ``audio–text–image–mask'' quadruples. Specifically, we employ an ambiguity filtering mechanism to ensure referential uniqueness, and incorporate eight distinct timbres alongside 4 realistic noises to simulate authentic environments. To drive the paradigm shift from text-centric to audio-centric, we propose the end-to-end Audio-centric Referring Expression Segmentation Network (ARES-Net). Distinct from discrete text-visual matching, our approach learns to establish a direct soft alignment between continuous acoustic manifolds and multi-scale visual representations. Comprehensive experiments show that our method obtains a peak accuracy of 64.23\% in oIoU and demonstrates robust generalization on a collected real-human test set. Furthermore, it significantly reduces total interaction latency to one-third of the text-based method. Our code is available at \url{https://github.com/LemonGauss/ARES-Net}.
\end{abstract}

\begin{figure}
  \begin{center}
    \centerline{\includegraphics[width=\columnwidth]{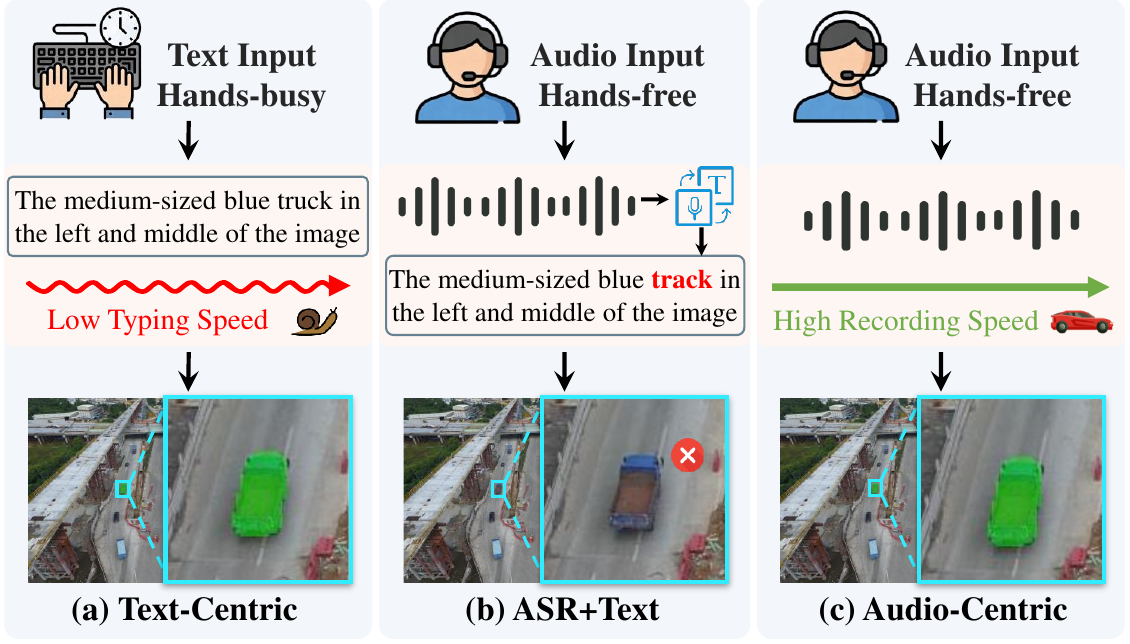}}
    \caption{Comparison of RES paradigms: (a) Text-Centric: inefficient manual typing; (b) ASR+Text: prone to transcription errors (e.g., ``truck" misrecognized as ``track") leading to failed segmentation; (c) Audio-Centric (Ours): robust, end-to-end, efficient perception without intermediate errors.}
    \label{fig:teaser}
  \end{center}
  \vskip -0.2in
\end{figure}

\section{Introduction}
\IEEEPARstart{R}{eferring} Expression Segmentation (RES)~\cite{xiao2025towards, 10227590} has increasingly evolved into UAV Referring Expression Segmentation (UAV-RES), building upon foundational UAV semantic segmentation benchmarks~\cite{lyu2020uavid, yao2025uemm}. As a pivotal perception task, UAV-RES aims to achieve pixel-level target localization in aerial imagery guided by natural language instructions~\cite{LI2025104817}. 
This capability is relevant to applications such as precision agriculture, disaster response, and traffic monitoring~\cite{LI2025104817, zhu2022detection}.

Recent advances in visual grounding and remote sensing perception involve both foundation models and task-specific architectures~\cite{chen2026evaluating, liu2025boost, BO202639, LU2026431, RemoteAgent}. For UAV-RES, these approaches leverage large-scale pre-training for visual-semantic alignment or specialized designs for aerial geometry and small targets. Both routes predominantly use text to specify the target. Manual text entry takes time and may divert visual attention from the task interface~\cite{wickens2002multiple}, as illustrated in Figure~\ref{fig:teaser}~(a). Speech provides a faster, hands-free input channel. A straightforward solution is to cascade Automatic Speech Recognition (ASR) with a text-based segmentation model, forming the ASR+Text pipeline in Figure~\ref{fig:teaser}~(b). However, noise and phonetic ambiguity can alter target categories, attributes, or spatial relations during transcription. These errors can propagate to mask prediction because the segmentation model receives the transcript rather than the original speech. ASR also adds a sequential processing stage, increasing system complexity and processing overhead.

Motivated by findings that spoken interaction achieves significantly higher throughput than typing~\cite{ruan2016speech}, we propose the Audio-Centric Referring Expression Segmentation (ARES) paradigm in Figure~\ref{fig:teaser}~(c). ARES advocates a fundamental shift in UAV perception from ``Text-Centric'' to ``Audio-Centric''. Distinct from existing approaches, ARES fundamentally reconstructs the cross-modal alignment pipeline without a text mediator to establish a direct mapping between continuous acoustic features and pixel-level visual features. This direct mapping brings two advantages: (1) ARES accelerates interaction speed by directly processing speech, thereby freeing the operator's hands and removing the intermediate transcription stage. (2) This paradigm outperforms the ASR+Text methods in robustness. Since audio embeddings operate on a continuous manifold~\cite{hsu2021hubert}, our approach enables a soft alignment mechanism. This mechanism maintains semantic coherence even under phonetic deformation, providing fault tolerance for users in real-world applications.

The proposed ARES paradigm necessitates dedicated audiovisual data, which is currently scarce in UAV scenarios. Consequently, we construct a benchmark named Aerial-AudioRES, featuring four core characteristics: Ambiguity Filtering, Acoustic Diversity, Noise Robustness, and Real-World Evaluation. To overcome the referential ambiguity caused by densely packed, identical targets typical of aerial views, we employ an ambiguity filtering mechanism to yield non-ambiguous targets, ensuring a unique referential mapping for both audio and text within the image. Furthermore, we synthesize large-scale speech using real-human reference audio of eight distinct timbres, encouraging the model to learn diverse and stable acoustic feature distributions. To evaluate model's robustness, we construct noisy variants corrupted by four specific environmental interferences: microphone, outdoor, indoor, and human noise. For Real-World Evaluation, the benchmark incorporates a real-human recording test set alongside clean synthetic data to assess the model's generalization across diverse pronunciation habits.

To implement the ARES paradigm, we specifically design the ARES-Net model. Text-based models rely on discrete token mapping, where incorrect mappings can cause semantic errors. In contrast, ARES-Net extracts audio inputs into continuous dense features via a HuBERT-base speech encoder. Fine-grained cross-modal alignment supports accurate referring segmentation~\cite{10413654}. ARES-Net directly aligns continuous acoustic features with multi-scale visual representations, bypassing intermediate ASR transcription. Evaluations on Aerial-AudioRES demonstrate that ARES-Net achieves state-of-the-art accuracy, achieving 43.81\% mIoU and 64.23\% oIoU. Furthermore, it exhibits strong robustness against real-human speech and complex noise, significantly enhancing interaction efficiency over manual text inputs.

In summary, this work makes the following contributions to the field of aerial vision-language understanding:
\begin{itemize}
\item We propose the \textbf{ARES} paradigm, an audio-centric referring segmentation task for aerial scenes, and design the \textbf{ARES-Net} as its specific implementation. To the best of our knowledge, this represents the first exploration of an audio-centric paradigm for UAV scenarios that establishes a direct and robust mapping between continuous acoustic waveforms and pixel-level visual features.

\item We construct the \textbf{Aerial-AudioRES} benchmark with an automatic data curation pipeline. It possesses rich acoustic diversity, encompassing real-human recordings and multiple noisy test sets, which facilitates comprehensive evaluation in realistic scenarios.

\item Experiments demonstrate ARES-Net's superior performance over text-based baselines and ASR-based approaches. It also demonstrates remarkable robustness against input distortions where other models fail.
\end{itemize}

\section{Related Work}
\subsection{Referring Expression Segmentation}

Referring Expression Segmentation (RES) is a fundamental task connecting visual and linguistic modalities. In the general domain, LAVT~\cite{yang2024language} and CrossVLT~\cite{10345690} improve referring segmentation through vision-language fusion. To further refine the alignment accuracy, FIANet~\cite{lei2024exploring} decouples referring expressions into object and spatial components, while CADFormer~\cite{liu2025cadformer} employs semantic mutual guidance to localize dense small targets. To accommodate the unique constraints of overhead imagery, specialized frameworks for aerial RES are proposed. RMSIN~\cite{liu2024rotated} introduces adaptive rotated convolutions, whereas AeroReformer~\cite{LI2025104817} utilizes multi-scale interaction to capture tiny targets. Additionally, while Vision-Language Models (VLMs) such as Falcon~\cite{yao2025falcon}, LHRS-Bot~\cite{muhtar2024lhrs}, EarthGPT~\cite{zhang2024earthgpt}, RemoteReasoner~\cite{yao2026remotereasoner}, and GeoChat~\cite{kuckreja2024geochat} have emerged for broad multimodal remote sensing comprehension, their heavy reliance on large language models limits parameter efficiency for dense visual perception. Foundation models such as RemoteSAM~\cite{yao2025remotesam} and Heterogeneous Knowledge Distillation~\cite{liu2026heterogeneous}, provide robust baselines for earth observation. IteRPrimE~\cite{wang2025iterprime} employs iterative Grad-CAM refinement and primary word emphasis to overcome positional insensitivity and manage complex semantic relations in out-of-domain scenarios. Benchmarks like RefDrone~\cite{sun2025refdrone} provide robust evaluation platforms for UAV scenarios. Despite these methods' successes, the inherent latency and input constraints of text-based interaction have become a primary bottleneck for overall interaction efficiency.

\subsection{Audio-Visual Modeling}

Building on vision-language models like CLIP~\cite{radford2021learning}, multi-modal alignments are widely explored. For instance, AudioCLIP~\cite{guzhov2022audioclip} performs global-level contrastive learning on general environmental audio, and Ref-AVS~\cite{wang2024ref} introduces audio-visual segmentation. However, AudioCLIP and Ref-AVS focuses on sound-emitting objects. In our ARES task, audio serves as external control instructions rather than inherent target properties. To achieve direct speech-visual interaction, models like SpeechCLIP~\cite{shih2023speechclip}, YOSS~\cite{10889085}, and RAGIS~\cite{santos2025layover} bypass text transcription. However, they share critical limitations: SpeechCLIP focuses on global-level alignment without pixel-level dense prediction; YOSS is restricted to simple word-level instructions; and RAGIS remains a preliminary evaluation. Crucially, none handle fine-grained segmentation driven by complex, sentence-level instructions, nor adapt to aerial domain constraints. Our work addresses this gap by establishing an end-to-end framework directly aligning continuous acoustic manifolds of complex commands with multi-scale visual features for UAV audio-centric referring expression segmentation.

\section{Theoretic Motivation}
\subsection{Task Formulation}

In the task of Audio-Centric Referring Expression Segmentation (ARES), we are given an aerial image $I \in \mathbb{R}^{3 \times H \times W}$ and a corresponding audio instruction $A \in \mathbb{R}^L$. ARES aims to learn a mapping function $\mathcal{F}: (I, A) \to M$ that produces a binary mask $M \in \{0, 1\}^{H \times W}$. Formally, the objective is to predict the pixel-wise probability $P(M_{i,j}=1 | I, A)$, determining whether the spatial coordinate $(i,j)$ belongs to the acoustically described entity. Unlike conventional text-based paradigms, ARES requires the model to locate visual targets directly using continuous acoustic features, rather than relying on discrete text tokens.

\subsection{Temporal Efficiency over Text}

We estimate input-time differences between spoken instructions and manual text entry, as illustrated in Figure~\ref{fig:teaser}~(a). We systematically quantify the speed advantage of audio over text from the temporal dimension. The user can generate a semantically corresponding audio instruction $A$ and a text instruction $T$, where $A$ and $T$ have the same length of words.
According to existing research~\cite{ruan2016speech}, human speech input speed (161.2 words per minute, WPM) is significantly faster than keyboard input speed (53.46, WPM). Therefore ARES enhances interaction efficiency via faster transmission while maintaining strict semantic consistency.

\subsection{Information Loss through ASR}

We analyze potential information loss caused by ASR transcription in the pipeline shown in Figure~\ref{fig:teaser}~(b). Mutual information $\mathcal{I}$~\cite{shannon1948mathematical} is evaluated conditional on the aerial image $I$. For a fixed deterministic ASR decoder $h(\cdot)$, the audio is converted into the transcript $\hat{T}=h(A)$. The text-based model $\mathcal{G}(\cdot)$ then predicts a segmentation mask $M_{\hat{T}}=\mathcal{G}(I,\hat{T})$ using the image and this transcript. The process defines the Markov chain~\cite{shannon1948mathematical} $X\to A\to\hat{T}\to M_{\hat{T}}$. The data processing inequality~\cite{beaudry2012intuitive} gives $\mathcal{I}(X;M_{\hat{T}}\mid I)\le\mathcal{I}(X;\hat{T}\mid I)\le\mathcal{I}(X;A\mid I)$, where $X$ represents the referring intent. Transcription may discard acoustic cues or introduce recognition errors that alter target descriptions. The relevant question is whether the transcript preserves the intent information available in the audio.

The direct model $\mathcal{F}(\cdot)$ predicts $M=\mathcal{F}(I,A)$, with $\mathcal{I}(X;M\mid I)\le\mathcal{I}(X;A\mid I)$. The difference between the information available in audio and its transcript satisfies:
\begin{equation}
\mathcal{I}(X;A\mid I)-\mathcal{I}(X;\hat{T}\mid I)
=\mathcal{I}(X;A\mid\hat{T},I)\ge 0.
\end{equation}
The residual term captures intent information retained in the audio but absent from its transcript. It therefore characterizes the information that direct speech conditioning can access before the final segmentation stage. A strict gap requires the remaining conditional mutual information to be positive. This motivates direct speech conditioning when transcription removes information needed for visual grounding. For synthetic samples, audio synthesis uses $T$ as its source caption. If synthesis variables carry no additional target information, then $\mathcal{I}(X;A\mid T,I)=0$. Audio therefore introduces no extra referring semantics beyond  $T$ under this assumption. ARES learns acoustic features directly under segmentation supervision, avoiding an intermediate ASR transcript.



\begin{figure*}[t]
  \begin{center}
    \centerline{\includegraphics[width=\textwidth]{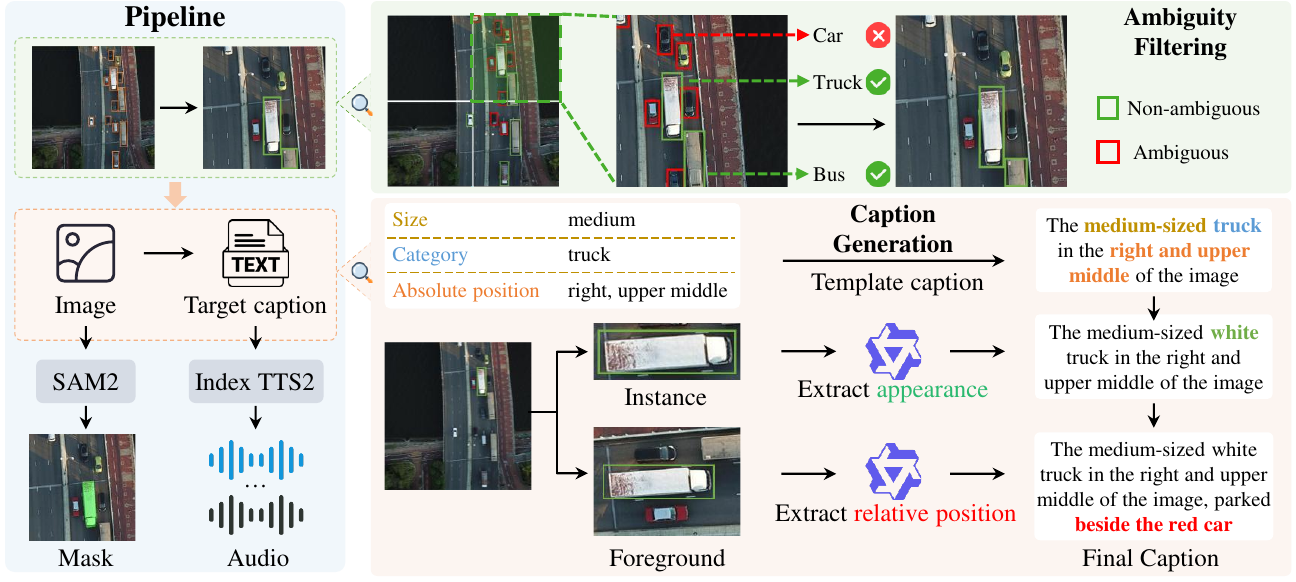}}
    \caption{Overview of the Aerial-AudioRES construction pipeline. It combines ambiguity-aware target selection and progressive caption generation with automated mask extraction and speech synthesis to construct aligned audio-text-image-mask quadruplets for aerial referring segmentation.}
    \label{fig:pipeline}
  \end{center}
  \vskip -0.2in
\end{figure*}

\section{The Aerial-AudioRES Benchmark}

\subsection{Data Generation Pipeline}

As illustrated in Figure~\ref{fig:pipeline}, we propose an automated dataset construction pipeline based on VisDrone imagery~\cite{zhu2022detection}. The pipeline employs an Ambiguity Filtering mechanism that retains targets whose category occurs only once within their spatial quadrant, establishing a controlled target set based on category and spatial location. For each retained target, a progressive caption generation strategy integrates statistical and visual attributes to produce a semantic description. Specifically, template captions are constructed from known statistical attributes, including size $\omega_{siz}$, category $\omega_{cls}$, and absolute position $\omega_{abs}$. These templates are enriched with appearance attributes $\omega_{app}$ extracted from instance images using Qwen2.5-VL-32B~\cite{bai2025qwen2}. The descriptions are further refined by prompting the same model to extract relative position attributes $\omega_{rel}$ from each target's foreground image, yielding the final captions. SAM2~\cite{ravi2024sam2} generates the corresponding target masks, while Index TTS 2~\cite{zhou2025indextts2} synthesizes audio instructions from the captions, conditioned on eight real-human reference recordings.
The post-generation quality control procedure is described in
Section~\ref{sec:quality_control}.

\subsection{Benchmark Characteristics}
To bridge the critical data gap in this domain, we present Aerial-AudioRES, the first benchmark specifically designed for audio-centric referring expression segmentation in aerial scenes. The synthetic subset comprises 28,460 ``audio-text-image-mask'' quadruplets. It is divided into training, validation, and test sets at an 8:1:1 ratio. A separate human-speech test set contains 189 recordings. 
The benchmark is constructed around four complementary components for evaluating speech-conditioned segmentation in aerial scenes: spatial filtering for controlled target selection, synthetic audio generation with eight reference voices via Index TTS 2~\cite{zhou2025indextts2}, controlled acoustic interference, and a dedicated human-speech test set. These features collectively establish a comprehensive evaluation paradigm, as detailed below.

\subsubsection{Ambiguity Filtering}
To ensure a deterministic mapping between instructions and targets and to address the semantic ambiguity caused by excessive similar targets in dense UAV perspectives, we introduce an ambiguity-aware filtering mechanism via spatial constraints. Specifically, we divide the original image into four quadrants. A candidate is defined as a qualified, non-ambiguous target only when it is the unique representative of its category within its quadrant. By retaining one instance per category in each quadrant, the filter establishes a controlled single-target evaluation setting. Category and quadrant information identify the candidate, while appearance attributes and relative relations enrich its description. This setup supports comparisons of speech-conditioned visual grounding under a shared target-selection protocol.

\subsubsection{Acoustic Diversity}
Index TTS 2~\cite{zhou2025indextts2} enables reference-conditioned speech synthesis with control over timbre and emotion, supporting varied acoustic realizations of the same instruction. By conditioning the synthesis on 8 selected real-human reference audios (3 males and 5 females), the generated commands precisely mimic human speech, naturally inheriting the significant variations in fundamental frequency trajectories and articulatory rhythms of the reference audios. The per-timbre experiment evaluates whether these acoustic variations affect segmentation. 

\subsubsection{Noise Robustness}
To simulate realistic physical environments, we introduce four distinct categories of acoustic interference. The microphone condition combines Brownian noise with a reverberation effect, alongside noises introduced from the ESC-50 dataset~\cite{piczak2015esc} such as outdoor noise (e.g., rain), human ambient chatter (e.g., coughing), and indoor noise (e.g., door knocking). All four noise types are systematically injected into the clean synthetic audio with controlled Signal-to-Noise Ratios (e.g., 15dB for ESC-50 noise and 10–20dB for dynamic microphone reverberation). This evaluation protocol rigorously tests the model's resilience against complex real-world acoustic degradation under diverse conditions.

\subsubsection{Real-World Evaluation}
Beyond synthetic data, we construct a dedicated real-human test set comprising a total of 189 recordings from 4 independent speakers. To accurately reflect the complexity of human-UAV interaction, these recordings are not merely scripted mechanical readings. They contain spontaneous speaking rates, natural accents, and distinct personal timbres. The human-speech test set evaluates transfer from synthetic instructions to human recordings. It captures speaker-dependent pronunciation, speaking rate, and prosodic variation while retaining the reference descriptions used to specify the visual targets. We additionally apply the same four noise conditions to the human recordings. This enables robustness comparisons across synthetic and human speech under a shared noise protocol.

\subsection{Annotation Quality Control}
\label{sec:quality_control}

We implement a post-generation quality control process to ensure the reliability of Aerial-AudioRES. For each object class, mean bounding-box areas are computed. Samples whose generated mask area was below 50\% of this class-specific mean were discarded. We manually reviewed a randomly selected 10\% of the quadruplets, comprising approximately 2,800 samples. The review assessed semantic alignment among captions, segmentation masks, and audio instructions. More than 95\% of the reviewed samples exhibited neither visual hallucinations nor substantial segmentation leakage. These results confirm that the Aerial-AudioRES benchmark provides high-quality, reliable supervision for model training in audio-driven UAV referring expression segmentation.

\section{Methodology}

\subsection{Model Framework}

\begin{figure}
  \begin{center}
        \centerline{\includegraphics[width=\columnwidth]{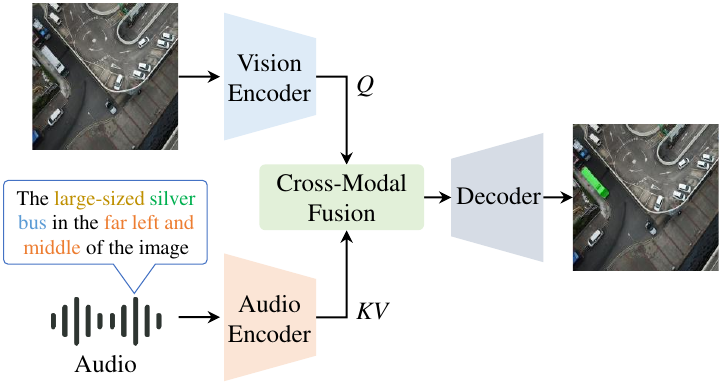}}
    \caption{Overview of the ARES-Net. The model aligns acoustic features with spatial visual representations and decodes the fused features into a target mask, enabling end-to-end prediction without intermediate ASR transcription.}
    \label{fig:method}
  \end{center}
  \vskip -0.2in
\end{figure}

ARES-Net implements direct speech-conditioned segmentation by adapting the RMSIN framework~\cite{liu2024rotated}. As illustrated in Figure~\ref{fig:method}, the model consists of visual encoding, acoustic encoding and refinement, cross-modal fusion, and mask prediction as follows:

\subsubsection{Vision Branch}

We retain the Swin-Transformer base backbone~\cite{liu2021swin} of RMSIN, using pre-trained visual representations~\cite{liu2021swin} to extract multi-scale features from $I$. At each scale, the visual features are flattened into a sequence of spatial tokens. Coarse-scale features provide scene context for locating the referred instance, while finer-scale features retain spatial detail for mask reconstruction.

\subsubsection{Audio Branch}

We employ pre-trained HuBERT-base~\cite{hsu2021hubert} to encode $A$ into frame-level acoustic features. HuBERT temporally downsamples the waveform into acoustic tokens. Spoken words can span multiple acoustic frames, motivating local refinement before visual alignment. We therefore place a single-layer Longformer block~\cite{Beltagy2020Longformer} after HuBERT-base and before cross-modal fusion. It uses eight attention heads and an attention window of $w=32$. Sliding-window attention aggregates neighboring acoustic tokens with complexity $\mathcal{O}(L_a w)$, where $L_a$ represents the number of tokens in the refined acoustic sequence. The refined output is $F_a\in\mathbb{R}^{L_a\times d_a}$, where $L_a$ is the acoustic token count and $d_a$ is the feature dimension. This sequence incorporates local acoustic context before spatially conditioned fusion.

\subsubsection{Cross-Modal Fusion}

We adapt the Pixel-Word Attention Module (PWAM) from LAVT~\cite{yang2024language} to use refined acoustic features in place of BERT-based text representations~\cite{devlin2019bert}. Learned projections place the two modalities in compatible attention spaces. Visual tokens provide the queries, while the refined acoustic features provide the keys and values. Cross-modal attention associates acoustic semantics with relevant visual regions~\cite{10812843}. The aggregated acoustic features are multiplied element-wise with projected visual features to produce speech-conditioned visual representations. The fused sequence retains the spatial organization of the visual branch.

\subsubsection{Decoder}

Following RMSIN~\cite{liu2024rotated}, the decoder combines features from multiple scales through top-down interpolation and concatenation. Coarse features provide semantic context, while finer features supply complementary spatial detail for mask reconstruction~\cite{10510606}. The merged representations estimate the foreground probability $p_{i,j}$ at each image pixel. These probabilities are converted into the final binary mask $M\in{0,1}^{H\times W}$, thereby completing the mapping $M=\mathcal{F}(I,A)$ defined in the task formulation.

\begin{table*}[t]
\caption{Performance on the Aerial-AudioRES benchmark (clean synthetic set). ASR+Text baselines default to Whisper-Base. Best and second-best results are \textbf{bold} and \underline{underlined}.}
\label{tab:main_results}
\begin{center}
    \resizebox{\textwidth}{!}{%
    \renewcommand{\arraystretch}{1.15} 
    \setlength{\aboverulesep}{0pt}     
    \setlength{\belowrulesep}{0pt}     
    \setlength{\tabcolsep}{6pt} 
    
    \begin{tabular}{ccccc|ccccccc}
    \toprule
    Method & Vision & Text/Audio & Publication & Modality & P@0.5 & P@0.6 & P@0.7 & P@0.8 & P@0.9 & mIoU & oIoU \\ 
    \midrule
    
    \rowcolor{gray!20}
    \multicolumn{12}{l}{\textit{Foundation Models}} \\
    
    \multirow{2}{*}{RemoteSAM~\cite{yao2025remotesam}} & \multirow{2}{*}{Swin-B} & \multirow{2}{*}{Bert-B} & \multirow{2}{*}{MM25} & Text & 6.57 & 5.87 & 5.02 & 4.04 & 2.53 & 6.63 & 1.12 \\
                                                       &                         &                         &                       & ASR+Text     & 6.32 & 5.66 & 4.85 & 3.83 & 2.28 & 6.39 & 1.02 \\ 
    \multirow{2}{*}{IteRPrime~\cite{wang2025iterprime}} & \multirow{2}{*}{ViT-B/16} & \multirow{2}{*}{Bert-B} & \multirow{2}{*}{AAAI25} & Text & 8.08 & 7.91 & 7.66 & 7.13 & 6.01 & 7.75 & 12.49 \\
                                                       &                         &                         &                         & ASR+Text     & 7.87 & 7.66 & 7.34 & 6.78 & 5.83 & 7.53 & 12.45 \\
    \multirow{2}{*}{Seg-Zero~\cite{liu2025seg}}  & \multirow{2}{*}{Qwen2.5-VL-7B} & \multirow{2}{*}{Qwen2.5-7B} & \multirow{2}{*}{ArXiv25} & Text & 33.03 & 31.17 & 29.06 & 26.07 & 21.64 & 30.86 & 32.90 \\
                                                 &                         &                         &                          & ASR+Text         & 32.68 & 30.71 & 28.95 & 25.69 & 21.22 & 28.24 & 30.41 \\
    
    \rowcolor{gray!20}
    \multicolumn{12}{l}{\textit{Specialized Models}} \\
    
    \multirow{2}{*}{AeroReformer~\cite{LI2025104817}} & \multirow{2}{*}{Swin-B} & \multirow{2}{*}{Bert-B} & \multirow{2}{*}{IJAEO25} & Text & 45.40 & 38.69 & 31.55 & 21.89 & 8.71 & 39.71 & 57.73 \\
                                                      &                         &                         &                          & ASR+Text        & 44.91 & 38.23 & 31.17 & 22.17 & 8.54 & 39.49 & 57.95 \\ 
    \multirow{2}{*}{CADFormer~\cite{liu2025cadformer}}    & \multirow{2}{*}{Swin-B} & \multirow{2}{*}{Bert-B} & \multirow{2}{*}{JSTARS25} & Text & 46.13 & 39.70 & 32.50 & 22.66 & 9.38 & 40.35 & 55.30 \\
                                                          &                         &                         &                           & ASR+Text       & 45.68 & 39.39 & 32.22 & 22.24 & 9.45 & 40.10 & 55.17 \\
    \multirow{2}{*}{FIANet~\cite{lei2024exploring}}       & \multirow{2}{*}{Swin-B} & \multirow{2}{*}{Bert-B} & \multirow{2}{*}{TGRS24}  & Text & 45.82 & 40.13 & 34.05 & \underline{25.37} & \underline{12.79} & 41.01 & 55.20 \\
                                                          &                         &                         &                          & ASR+Text        & 45.47 & 39.74 & 33.31 & 25.26 & 12.58 & 40.58 & 54.60 \\
    \multirow{2}{*}{RefSegformer~\cite{wu2024toward}} & \multirow{2}{*}{Swin-B} & \multirow{2}{*}{Bert-B} & \multirow{2}{*}{TIP24}   & Text & 43.96 & 38.83 & 32.19 & 23.83 & 12.72 & 39.10 & \underline{59.16} \\
                                                      &                         &                         &                          & ASR+Text        & 43.78 & 38.30 & 32.29 & 23.54 & 12.58 & 38.97 & 58.97 \\ 
    \multirow{2}{*}{RMSIN~\cite{liu2024rotated}}        & \multirow{2}{*}{Swin-B} & \multirow{2}{*}{Bert-B} & \multirow{2}{*}{CVPR24}  & Text & \underline{48.24} & \underline{41.95} & \underline{34.57} & 24.53 & 12.12 & \underline{42.38} & 58.73 \\
                                                        &                         &                         &                          & ASR+Text        & 47.19 & 41.29 & 34.29 & 24.56 & 11.70 & 41.74 & 58.33 \\
    
    \midrule[0.15pt]
    \rowcolor{blue!10} 
    \textbf{ARES-Net(Ours)} \rule[-1.2ex]{0pt}{4ex} & \textbf{Swin-B} & \textbf{Hubert-B} & - & \textbf{Audio} & \textbf{49.09} & \textbf{43.43} & \textbf{36.89} & \textbf{27.16} & \textbf{14.05} & \textbf{43.81} & \textbf{64.23} \\
    \bottomrule
    \end{tabular}%
    }
\end{center}
\end{table*}

\subsection{Training Objective}

We jointly fine-tune all layers of the vision encoder, HuBERT-base, and decoder under segmentation supervision. This allows the segmentation objective to guide acoustic feature adaptation through cross-modal fusion and mask prediction. Training combines weighted cross-entropy loss $\mathcal{L}_{wce}$~\cite{xie2015holistically} and the dice loss $\mathcal{L}_{dice}$~\cite{milletari2016v} to account for foreground-background imbalance and optimize target-mask overlap. $\mathcal{L}_{wce}$ mitigates the class imbalance issue by introducing balancing coefficients for different categories:
\begin{equation}
\begin{aligned}
\mathcal{L}_{wce}
=-\sum_{i=1}^{H}\sum_{j=1}^{W}\big[
&w_1y_{i,j}\log p_{i,j}\\
&+w_0(1-y_{i,j})\log(1-p_{i,j})\big],
\end{aligned}
\end{equation}
where $y_{i,j}\in\{0,1\}$ is the ground-truth target label and $p_{i,j}$ is the predicted foreground probability. The coefficients $w_1$ and $w_0$ balance foreground and background contributions, respectively. $\mathcal{L}_{dice}$ encourages overlap between the probability map and the ground-truth mask:
\begin{equation}
\mathcal{L}_{dice}
=1-\frac{2\sum_{i=1}^{H}\sum_{j=1}^{W}y_{i,j}p_{i,j}}
{\sum_{i=1}^{H}\sum_{j=1}^{W}y_{i,j}
+\sum_{i=1}^{H}\sum_{j=1}^{W}p_{i,j}}.
\end{equation}
Weighted cross-entropy supplies pixel-wise supervision, while dice couples predictions across the mask through a global overlap ratio. Their combination trains local foreground decisions together with the overall target extent. The total training objective is:
\begin{equation}
\mathcal{L}_{total}
=(1-\lambda)\mathcal{L}_{wce}
+\lambda\mathcal{L}_{dice},
\end{equation}
where $\lambda$ controls the relative contribution of the two losses.

\section{Experiments}

\subsection{Implementation Details}
\subsubsection{Training Settings}

Our experiments were conducted using two NVIDIA GeForce RTX 4090 GPUs, which served as the unified platform for both training and latency measurements. To guarantee fair comparison, all baselines were evaluated under identical dataset splits and augmentations. Detailed architectures and parameter counts are in the Appendix A. Input images were resized to a fixed resolution of $480 \times 480$. The training process spanned 60 epochs, utilizing the AdamW optimizer with an initialized learning rate of $3 \times 10^{-5}$. Gradient accumulation with steps set to 4 was applied. The trade-off hyperparameter $\lambda$ was empirically set to 0.1. Whisper-Base~\cite{radford2023robust} is employed as our ASR model.

\subsubsection{Evaluation Metrics} 

For a more comprehensive and rigorous performance evaluation of our proposed method, we utilize Precision@$k$, mean Intersection over Union (mIoU), and Overall IoU (oIoU). Precision@$k$, evaluated at multiple predefined thresholds $k \in \{0.5, 0.6, 0.7, 0.8, 0.9\}$, measures the percentage of samples where the resulting IoU exceeds $k$. mIoU calculates the average IoU score across all available test samples to quantitatively evaluate the overall instance-level segmentation quality, while oIoU provides a comprehensive global assessment of pixel-level alignment accuracy.

\subsection{Main Results}
\subsubsection{Performance on Aerial-AudioRES}
We validated the segmentation performance of ARES-Net on the Aerial-AudioRES clean synthetic test set. To demonstrate its superior capabilities, we systematically compared our audio-centric approach against several existing text-based SOTA methods. All specialized models utilize the Swin-base and BERT-base-uncased backbones. For these baseline methods, we evaluated two distinct input settings. The Text setting uses the original generated text that was utilized to synthesize the audio. The ASR (Whisper-Base)+Text setting uses the text transcribed directly from the audio input. As shown in Table~\ref{tab:main_results}, the ASR+Text performance is consistently lower mIoU than the Text performance across all models because the speech recognition process introduces unavoidable transcription errors and information loss. ARES-Net achieves the highest mIoU and oIoU among the evaluated methods, reaching 43.81\% and 64.23\%, respectively. Crucially, because ARES-Net shares the identical vision encoder, fusion module, and decoder with the text-based SOTA model RMSIN, our comparison serves as a strict modality ablation. The absolute improvements of 1.43\% in mIoU and 5.5\% in oIoU over the Text setting of RMSIN thus strictly stem from the modality shift. The results demonstrate that learning directly from continuous audio signals yields more precise segmentation performance compared to existing text-based methods.

\begin{table}[t!]
    \centering
    \caption{Evaluation on the clean real-human test set. ASR defaults to Whisper-Base.}
    \label{tab:real_human}
    \begin{small}
        \renewcommand{\arraystretch}{1.15} 
        
        \begin{tabularx}{\columnwidth}{c *{5}{>{\centering\arraybackslash}X}}
        \toprule
        Method & Modality & P@0.5 & P@0.9 & mIoU & oIoU \\
        \midrule
        RMSIN & ASR+Text & 38.10 & 4.76 & 32.79 & 43.80 \\
        
        \rowcolor{blue!10} \textbf{ARES-Net} & \textbf{Audio} & \textbf{46.56} & \textbf{16.93} & \textbf{41.65} & \textbf{64.64} \\
        \bottomrule
        \end{tabularx}
    \end{small}
    \vskip -0.1in
\end{table}

\begin{figure}[t!]
  \begin{center}
    \centerline{\includegraphics[width=\columnwidth]{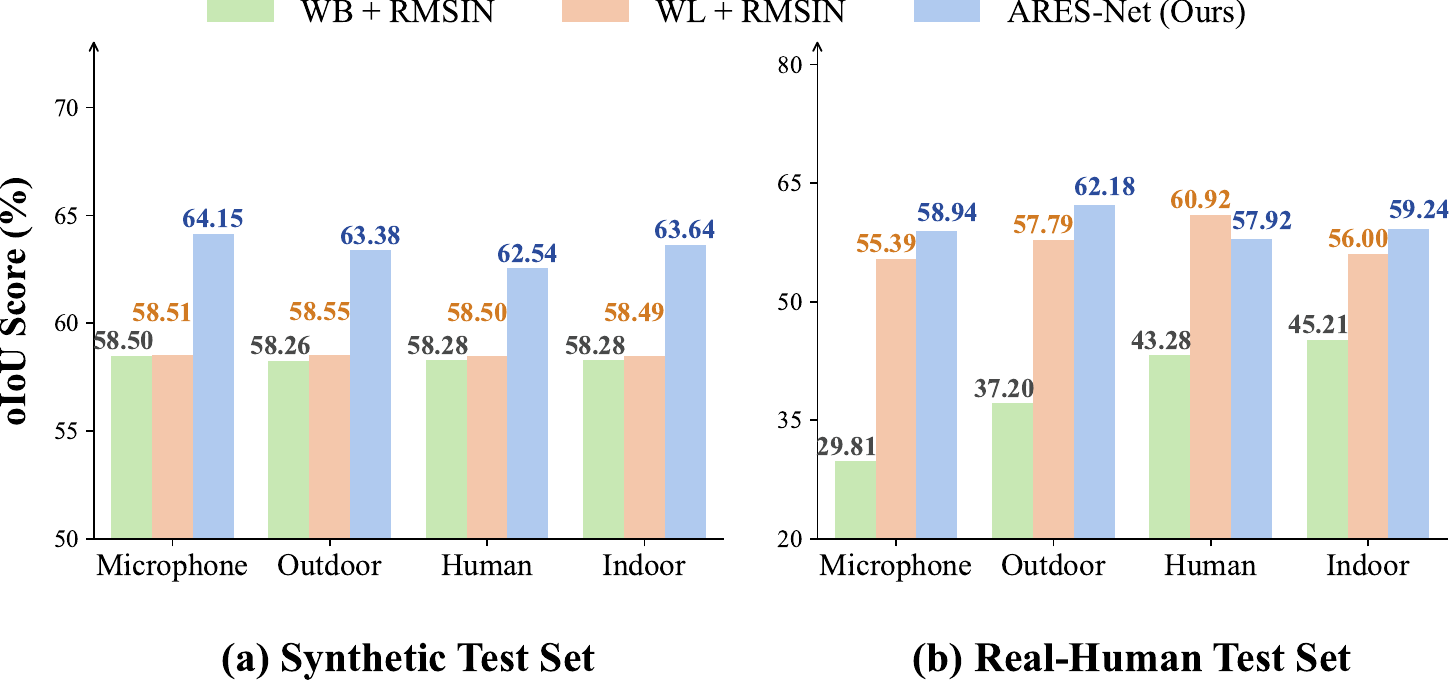}}
    \caption{Performance comparison of ARES-Net and ASR + RMSIN across various noisy test sets. ``WB'' and ``WL'' denote Whisper-Base and Whisper-Large-V3, respectively.}
    \label{fig:noise}
  \end{center}
  \vskip -0.2in
\end{figure}

\subsubsection{Generalization to Real-human Speech}
This experiment examines the generalization ability of the model towards real recording inputs using a test set comprising 189 real-human recordings. This setup reflects system performance when encountering non-ideal pronunciation habits. Results in Table \ref{tab:real_human} indicate that ARES-Net achieves a leap of 8.86 in mIoU and 20.84\% in oIoU compared to the ASR+Text scheme. ASR+Text passes recognized words to the segmentation model, so errors in target descriptions can affect visual grounding. ARES-Net instead conditions mask prediction on HuBERT acoustic features~\cite{hsu2021hubert}, avoiding an intermediate transcript.


\subsection{Further Analysis}


\subsubsection{Robustness in Noisy Environments}

We compared ARES-Net with an ASR+Text baseline to evaluate robustness against noisy audio. The baseline pairs Whisper-Base~\cite{radford2023robust} with RMSIN, and both segmentation models use clean synthetic data for task-specific training. ARES-Net, equipped with a 94M HuBERT-base encoder, consistently outperforms the Whisper-Base cascade across the tested noisy conditions (Figure~\ref{fig:noise}). This advantage persists in most scenarios even when the ASR frontend is upgraded to the 1.55B-parameter Whisper-Large-V3. On real-human recordings, Whisper-Base exhibits WER above 34\% across the tested conditions, reaching 56.95\% under microphone noise (Table~\ref{tab:WER}). Recognition errors can alter target categories or spatial descriptions, making the referring cues supplied to RMSIN less reliable. ARES-Net bypasses this transcription stage and uses cross-modal attention to associate acoustic features with visual regions. This direct alignment allows acoustic evidence and visual context to jointly support target localization under noisy conditions.


\begin{figure*}[t]
  \begin{center}
    \centerline{\includegraphics[width=\textwidth]{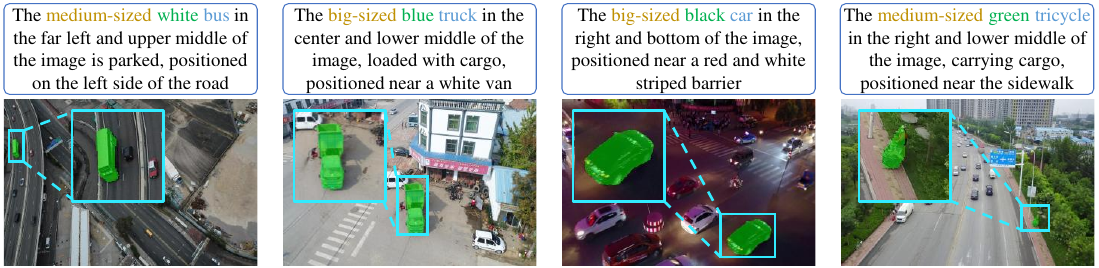}}
    \caption{Visualization of ARES-Net predictions on the noisy synthetic test set.}
    \label{fig:keshihua2}
  \end{center}
  \vskip -0.3in
\end{figure*}

\begin{table}[t!]
\caption{Comparison of WER \% across different noise scenarios on synthetic and real-human test sets.} 
\vskip -0.1in
\label{tab:WER}
\begin{center}
\begin{small}
    \setlength{\tabcolsep}{0pt} 
    \begin{tabular*}{\columnwidth}{@{\extracolsep{\fill}}cccccc}
    \toprule
    Dataset (WER $\downarrow$) & Clean & Microphone & Outdoor & Human & Indoor \\
    \midrule
    Synthetic & 4.22 & 5.50 & 5.72 & 5.98 & 5.53 \\
    Real-Human & 34.82 & 56.95 & 41.05 & 42.98 & 39.99 \\
    \bottomrule
    \end{tabular*}
\end{small}
\end{center}
\end{table}

\begin{table}
\caption{Comparison of interaction latency and parameter efficiency. ``WB'' and ``WL'' denote Whisper-Base and Whisper-Large-V3, respectively. ``Inf.'' stands for Inference Time.}
\vskip -0.1in
\label{tab:latency_comparison}
\begin{center}
\begin{small}
    \setlength{\tabcolsep}{1.5pt} 
    
    \begin{tabularx}{\columnwidth}{c >{\centering\arraybackslash}X c *{5}{>{\centering\arraybackslash}X}}
\toprule
    \multirow{2}{*}{Method} & \multicolumn{3}{c}{Params (M)} & \multicolumn{4}{c}{Latency (s)} \\
    \cmidrule(lr){2-4} \cmidrule(lr){5-8}
    & ASR & \mbox{RES Model} & Total & Input & ASR & Inf. & Total \\
\midrule
    RMSIN & -- & 240 & 240& 26.30 & -- & 0.03 & 26.33 \\
    WB+RMSIN & 74 & 240 & 314 & 8.72 & 0.26 & 0.03 & 9.01 \\
    WL+RMSIN & 1550 & 240 & 1790 & 8.72 & 0.76 & 0.03 & 9.51 \\
    
    \rowcolor{blue!10} \textbf{ARES-Net} & -- & 249 & 249 & 8.72 & -- & 0.13 & 8.85 \\
\bottomrule
\end{tabularx}
\end{small}
\end{center}
\vskip -0.2in
\end{table}

\subsubsection{Interaction and Model Efficiency}
To demonstrate the efficiency of directly using audio instructions to drive RES, we compared the latency and parameter scale of four methods in Table \ref{tab:latency_comparison}. Our evaluation focuses on the entire human-machine interaction cycle. According to the research~\cite{ruan2016speech}, human speech input speed (161.2 words per minute, WPM) is significantly faster than keyboard input speed (53.46 WPM). We calculate input time based on average instruction word counts. Because ARES-Net and RMSIN share the same architecture except for the audio/text encoder, and the audio tensor is dozens of times longer than the text tensor, the pure Inference Time of ARES-Net is slightly higher than that of RMSIN. While adopting Whisper-Large-v3 marginally improves cascaded accuracy, it incurs a 1.55B parameter burden(vs. 94M for HuBERT-base) and inflates ASR latency to 0.76s. In contrast, ARES-Net achieves the lowest Total Time without such massive memory overhead, resulting from bypassing the slow textual input process and intermediate ASR stages. Specifically, ARES-Net achieved an average total latency of 8.85s, reducing the overall task completion time to approximately $1/3$ of RMSIN.

\subsection{Ablation Studies}
\subsubsection{Effect of the Longformer Module}

We ablated Longformer to assess the contribution of local sequence refinement to audio-driven segmentation. RMSIN serves as a text reference to examine whether the benefit depends on the input representation. Table~\ref{tab:longformer_ablation} reports the four configurations on the clean synthetic test set. For ARES-Net, Longformer improves mIoU and oIoU by 1.96 and 2.84 percentage points, respectively. The results demonstrate that Longformer can improve visual grounding by aggregating linguistic evidence across neighboring HuBERT frames.
The text setting shows a smaller and mixed response to the same module. Adding Longformer to RMSIN increases oIoU by 0.38 percentage points but decreases mIoU by 0.12 points. BERT represents written instructions through contextualized lexical tokens, forming a more compact sequence of linguistic units. Additional local aggregation may therefore provide limited complementary information in this setting. These results suggest that the benefit of Longformer depends on the structure of the encoded sequence. The larger gains for audio support its use for refining dense acoustic representations in ARES-Net.

\begin{table}[t]
  \centering
  \caption{Longformer ablation on the clean synthetic test set. LF denotes Longformer.}
  \label{tab:longformer_ablation}
  \normalsize
  \setlength{\tabcolsep}{0pt}
  \renewcommand{\arraystretch}{1.1}
  \begin{tabular}{@{}
    >{\centering\arraybackslash}p{0.14\columnwidth}
    >{\centering\arraybackslash}p{0.22\columnwidth}
    >{\centering\arraybackslash}p{0.08\columnwidth}
    *{4}{>{\centering\arraybackslash}p{0.14\columnwidth}}
    @{}}
    \toprule
    Input & Model & LF & P@0.5 & P@0.9 & mIoU & oIoU \\
    \midrule
    Text  & RMSIN    & w/o & \underline{48.24} & 12.12
          & \underline{42.38} & 58.73 \\
    Text  & RMSIN    & w/  & 47.58 & 11.28 & 42.26 & 59.11 \\
    Audio & ARES-Net & w/o & 47.79 & \underline{13.39}
          & 41.85 & \underline{61.39} \\
    \rowcolor{blue!10} \textbf{Audio} & \textbf{ARES-Net} & \textbf{w/}
          & \textbf{49.09} & \textbf{14.05}
          & \textbf{43.81} & \textbf{64.23} \\
    \bottomrule
  \end{tabular}
\end{table}

\begin{table}[t]
  \centering
  \caption{Ablation of HuBERT-base training strategies on the clean synthetic test set.}
  \label{tab:hubert_ablation}
  \normalsize
  \setlength{\tabcolsep}{0pt}
  \renewcommand{\arraystretch}{1.1}
  \begin{tabular}{@{}
    >{\raggedright\arraybackslash}p{0.32\columnwidth}
    *{4}{>{\centering\arraybackslash}p{0.17\columnwidth}}
    @{}}
    \toprule
    Strategy & P@0.5 & P@0.9 & mIoU & oIoU \\
    \midrule
    Frozen & 45.75 & \underline{12.12} & 40.61 & 60.04 \\
    LoRA (Attention) & \underline{47.33} & 11.91
                     & \underline{42.20} & \underline{60.38} \\
    \rowcolor{blue!10} Full fine-tuning & \textbf{49.09} & \textbf{14.05}
                     & \textbf{43.81} & \textbf{64.23} \\
    \bottomrule
  \end{tabular}
\end{table}

\subsubsection{Ablation Study on Audio Encoder Training }
To investigate the impact of the audio encoder's optimization on performance, we conducted ablation studies on the HuBERT module using the clean synthetic test set. We compared three settings: (1) \textbf{Frozen}: keeping the pre-trained parameters fixed; (2) \textbf{LoRA (Attention)}: applying Low-Rank Adaptation to the attention layers for efficient tuning; and (3) \textbf{Full Training}: unfreezing all parameters for holistic optimization.
As shown in Table~\ref{tab:hubert_ablation}, the full training strategy consistently achieves the highest performance across all metrics, significantly outperforming the Frozen baseline (e.g., +3.2\% in mIoU). This superiority naturally stems from the inherently high acoustic diversity of our dataset, which contains instructions with 8 distinct timbres. Full parameter training allows the encoder to flexibly and thoroughly adapt its feature space to these varied acoustic styles, capturing the unique spectral characteristics of each timbre better than partial fine-tuning strategies.

\begin{figure*}[t]
  \begin{center}
    \centerline{\includegraphics[width=\textwidth]{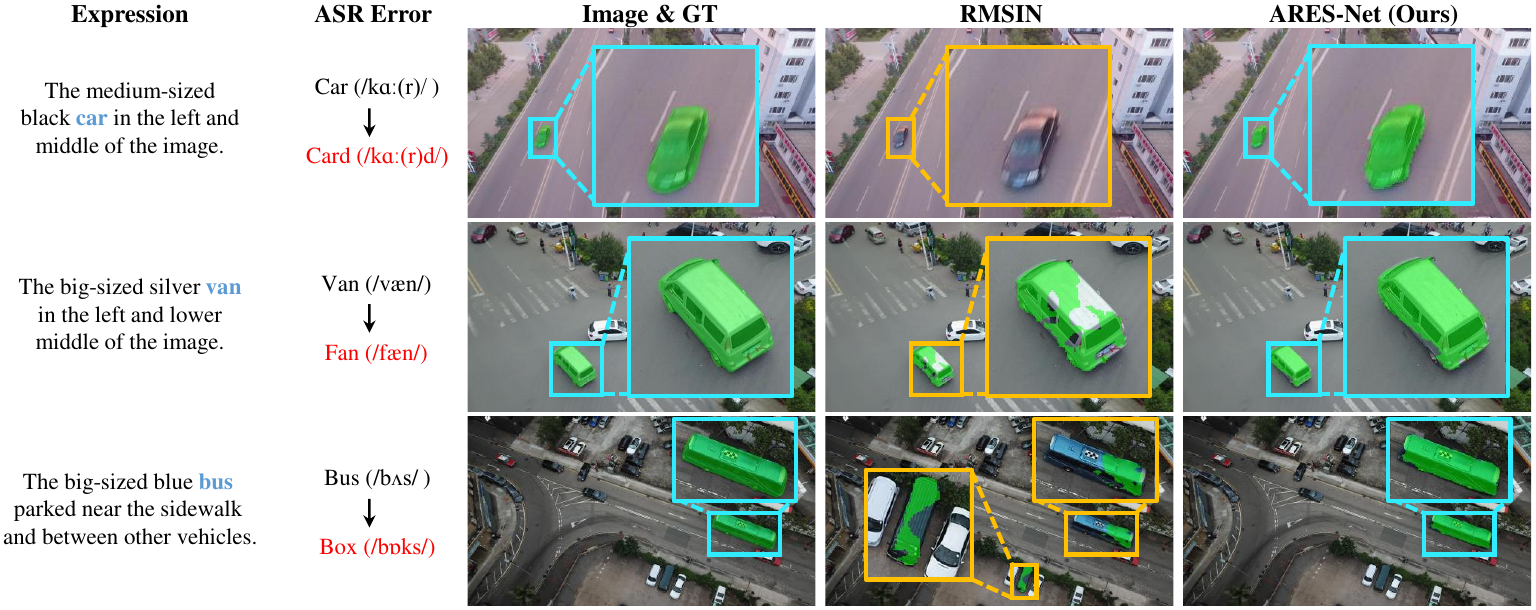}}
    \caption{Robustness against phonetic ambiguity. While phonetically similar words cause ASR transcription errors and RMSIN failure, ARES-Net successfully segments the query image.}
    \label{fig:keshihua}
  \end{center}
  \vskip -0.2in
\end{figure*}

\begin{table}
\caption{Impact of individual semantic components within instructions. Each configuration evaluates the addition of a single semantic attribute to the base target category.}
\label{tab:ablation_semantic}
\begin{center}
\begin{small}
\setlength{\tabcolsep}{3pt} 
\begin{tabularx}{\columnwidth}{c *{4}{>{\centering\arraybackslash}X}}
\toprule
Semantic Composition & P@0.5 & P@0.9 & mIoU & oIoU \\
\midrule
Base Category & 1.97 & 0.14 & 2.26 & 4.17 \\
+ Size & 10.96 & 2.57 & 10.74 & 24.77 \\
+ Appearance & 16.37 & 5.31 & 15.06 & 27.86 \\
+ Absolute Position & \underline{37.42} & \underline{6.64} & \underline{33.78} & \underline{54.42} \\
+ Relative Position & 8.36 & 1.72 & 7.93 & 12.43 \\
\midrule
\rowcolor{blue!10} \textbf{Full Expression (All)} & \textbf{49.09}  & \textbf{14.05} & \textbf{43.81} & \textbf{64.23} \\
\bottomrule
\end{tabularx}
\end{small}
\end{center}
\vskip -0.1in
\end{table}

\subsubsection{Semantic Unit Complexity Analysis}
We conducted ablation studies to validate the impact of the completeness of semantic units $\Omega$ on segmentation accuracy. As shown in Table~\ref{tab:ablation_semantic}, instructions containing solely category information $\omega_{cls}$ yield a meager mIoU of 2.26\%, demonstrating that sparse instructions lead to severe referential ambiguity within dense UAV scenes. Upon the respective introduction of appearance attributes $\omega_{attr}$ and absolute locations $\omega_{abs}$, model performance improved significantly, with mIoU surging to 15.06\% and 33.78\%, respectively. These gains show that appearance and absolute-position information provide useful cues for target localization. 
Category and absolute position units provide strong localization cues under the benchmark's quadrant filtering rule. Full expressions further improve segmentation by combining these units with additional attributes and relations. These results show that the type and combination of semantic units influence grounding performance. Integrating complementary units therefore strengthens audio-driven localization by jointly specifying target attributes and spatial context on this benchmark.

\subsection{Visualization Analysis}

We qualitatively examine target localization under acoustic interference. The predictions in Figure~\ref{fig:keshihua2} localize targets at different spatial scales and distinguish them from surrounding aerial backgrounds despite noise in the audio instructions. Spatial context provides complementary cues for associating the noisy instruction with a specific object in the scene.

We further compare ARES-Net with ASR + RMSIN under phonetic ambiguity, as shown in Figure~\ref{fig:keshihua}. In the illustrated cases, near-homophones such as ``Car'' and ``Card'' produce incorrect ASR target descriptions, leading RMSIN to segment unintended objects. A category substitution changes a central referring cue supplied to the segmentation model, even when the image remains unchanged. ARES-Net directly aligns the original acoustic features with visual evidence and identifies the intended targets. These examples demonstrate accurate localization under the tested acoustic conditions, despite training exclusively on clean synthetic data.


\section{Conclusion}
In this study, we propose the ARES paradigm, establishing an audio-centric perception foundation for referring expression segmentation in UAV scenarios. To realize this novel paradigm, we specifically design the ARES-Net model. By constructing an automated data pipeline, we developed the Aerial-AudioRES benchmark with 28,460 high-quality quadruplets, featuring multi-accent synthetic and real-human recordings. Extensive experiments demonstrate that ARES-Net achieves state-of-the-art accuracy on both mIoU and oIoU. Notably, compared to text-based methods, our framework accelerates interaction, cutting total latency to approximately one-third. Overall, direct audio perception serves as a more efficient and robust alternative to traditional text-based interaction for RES.

\bibliographystyle{IEEEtran}
\bibliography{ref}

@article{LI2025104817,
title={AeroReformer: Aerial referring transformer for UAV-based referring image segmentation},
  author={Li, Rui and Zhao, Xiaowei},
  journal={International Journal of Applied Earth Observation and Geoinformation},
  volume={143},
  pages={104817},
  year={2025},
  publisher={Elsevier}
}

@inproceedings{10889085,
  title={You only speak once to see},
  author={Yang, Wenhao and Wei, Jianguo and Lu, Wenhuan and Li, Lei},
  booktitle={ICASSP 2025-2025 IEEE International Conference on Acoustics, Speech and Signal Processing (ICASSP)},
  pages={1--5},
  year={2025},
  organization={IEEE}
}

@article{wu2024toward,
  title={Toward robust referring image segmentation},
  author={Wu, Jianzong and Li, Xiangtai and Li, Xia and Ding, Henghui and Tong, Yunhai and Tao, Dacheng},
  journal={IEEE Transactions on Image Processing},
  volume={33},
  pages={1782--1794},
  year={2024},
  publisher={IEEE}
}

@article{liu2025cadformer,
  title={Cadformer: Fine-grained cross-modal alignment and decoding transformer for referring remote sensing image segmentation},
  author={Liu, Maofu and Jiang, Xin and Zhang, Xiaokang},
  journal={IEEE Journal of Selected Topics in Applied Earth Observations and Remote Sensing},
  year={2025},
  publisher={IEEE}
}

@inproceedings{yao2025remotesam,
  title={Remotesam: Towards segment anything for earth observation},
  author={Yao, Liang and Liu, Fan and Chen, Delong and Zhang, Chuanyi and Wang, Yijun and Chen, Ziyun and Xu, Wei and Di, Shimin and Zheng, Yuhui},
  booktitle={Proceedings of the 33rd ACM International Conference on Multimedia},
  pages={3027--3036},
  year={2025}
}

@inproceedings{guzhov2022audioclip,
  title={Audioclip: Extending clip to image, text and audio},
  author={Guzhov, Andrey and Raue, Federico and Hees, J{\"o}rn and Dengel, Andreas},
  booktitle={ICASSP 2022-2022 IEEE International Conference on Acoustics, Speech and Signal Processing (ICASSP)},
  pages={976--980},
  year={2022},
  organization={IEEE}
}

@article{wickens2002multiple,
  title={Multiple resources and performance prediction},
  author={Wickens, Christopher D},
  journal={Theoretical issues in ergonomics science},
  volume={3},
  number={2},
  pages={159--177},
  year={2002},
  publisher={Taylor \& Francis}
}

@article{lei2024exploring,
  author  = {S. Lei and X. Xiao and T. Zhang and H.-C. Li and Z. Shi and Q. Zhu},
  title   = {Exploring fine-grained image-text alignment for referring remote sensing image segmentation},
  journal = {IEEE TGRS},
  year    = {2024}
}

@inproceedings{liu2024rotated,
  title={Rotated multi-scale interaction network for referring remote sensing image segmentation},
  author={Liu, Sihan and Ma, Yiwei and Zhang, Xiaoqing and Wang, Haowei and Ji, Jiayi and Sun, Xiaoshuai and Ji, Rongrong},
  booktitle={Proceedings of the IEEE/CVF Conference on Computer Vision and Pattern Recognition},
  pages={26658--26668},
  year={2024}
}

@article{yang2024language,
  title={Language-aware vision transformer for referring segmentation},
  author={Yang, Zhao and Wang, Jiaqi and Ye, Xubing and Tang, Yansong and Chen, Kai and Zhao, Hengshuang and Torr, Philip HS},
  journal={IEEE Transactions on Pattern Analysis and Machine Intelligence},
  volume={47},
  number={7},
  pages={5238--5255},
  year={2024},
  publisher={IEEE}
}

@article{sun2025refdrone,
  title={Refdrone: A challenging benchmark for referring expression comprehension in drone scenes},
  author={Sun, Zhichao and Liu, Yepeng and Su, Zhiling and Zhu, Huachao and Gu, Yuliang and Zou, Yuda and Liu, Zelong and Xia, Gui-Song and Du, Bo and Xu, Yongchao},
  journal={arXiv preprint arXiv:2502.00392},
  year={2025}
}

@article{santos2025layover,
  title={Layover or Direct Flight: Rethinking Audio-Guided Image Segmentation},
  author={Santos, Joel Alberto and Wu, Zongwei and Alameda-Pineda, Xavier and Timofte, Radu},
  journal={arXiv preprint arXiv:2511.22025},
  year={2025}
}

@inproceedings{shih2023speechclip,
  title={Speechclip: Integrating speech with pre-trained vision and language model},
  author={Shih, Yi-Jen and Wang, Hsuan-Fu and Chang, Heng-Jui and Berry, Layne and Lee, Hung-yi and Harwath, David},
  booktitle={2022 IEEE Spoken Language Technology Workshop (SLT)},
  pages={715--722},
  year={2023},
  organization={IEEE}
}

@article{ruan2016speech,
  title={Speech is 3x faster than typing for english and mandarin text entry on mobile devices},
  author={Ruan, Sherry and Wobbrock, Jacob O and Liou, Kenny and Ng, Andrew and Landay, James},
  journal={arXiv preprint arXiv:1608.07323},
  year={2016}
}

@inproceedings{zhou2025indextts2,
  title={Indextts2: A breakthrough in emotionally expressive and duration-controlled auto-regressive zero-shot text-to-speech},
  author={Zhou, Siyi and Zhou, Yiquan and He, Yi and Zhou, Xun and Wang, Jinchao and Deng, Wei and Shu, Jingchen},
  booktitle={Proceedings of the AAAI Conference on Artificial Intelligence},
  volume={40},
  number={41},
  pages={35139--35148},
  year={2026}
}

@misc{bai2025qwen2,
      title={Qwen2.5-VL Technical Report}, 
      author={Shuai Bai and Keqin Chen and Xuejing Liu and Jialin Wang and Wenbin Ge and Sibo Song and Kai Dang and Peng Wang and Shijie Wang and Jun Tang and Humen Zhong and Yuanzhi Zhu and Mingkun Yang and Zhaohai Li and Jianqiang Wan and Pengfei Wang and Wei Ding and Zheren Fu and Yiheng Xu and Jiabo Ye and Xi Zhang and Tianbao Xie and Zesen Cheng and Hang Zhang and Zhibo Yang and Haiyang Xu and Junyang Lin},
      year={2025},
      eprint={2502.13923},
      archivePrefix={arXiv},
      primaryClass={cs.CV},
      url={https://arxiv.org/abs/2502.13923}, 
}

@article{ravi2024sam2,
  author={Nikhila Ravi and Valentin Gabeur and Yuan-Ting Hu and Ronghang Hu and Chaitanya Ryali and Tengyu Ma and Haitham Khedr and Roman Rädle and Chloe Rolland and Laura Gustafson and Eric Mintun and Junting Pan and Kalyan Vasudev Alwala and Nicolas Carion and Chao-Yuan Wu and Ross Girshick and Piotr Dollár and Christoph Feichtenhofer},
  title   = {SAM 2: Segment Anything in Images and Videos},
  journal = {arXiv preprint arXiv:2408.00714},
  year    = {2024}
}

@article{beaudry2012intuitive,
  author={Normand J. Beaudry and Renato Renner},
  title   = {An intuitive proof of the data processing inequality},
  journal = {Quantum Information \& Computation},
  year    = {2012}
}

@article{shannon1948mathematical,
  title={A mathematical theory of communication},
  author={Shannon, Claude Elwood},
  journal={The Bell system technical journal},
  volume={27},
  number={3},
  pages={379--423},
  year={1948},
  publisher={Nokia Bell Labs}
}

@article{liu2025seg,
  title={Seg-zero: Reasoning-chain guided segmentation via cognitive reinforcement},
  author={Liu, Yuqi and Peng, Bohao and Zhong, Zhisheng and Yue, Zihao and Lu, Fanbin and Yu, Bei and Jia, Jiaya},
  journal={arXiv preprint arXiv:2503.06520},
  year={2025}
}

@inproceedings{wang2025iterprime,
  title={Iterprime: Zero-shot referring image segmentation with iterative grad-cam refinement and primary word emphasis},
  author={Wang, Yuji and Ni, Jingchen and Liu, Yong and Yuan, Chun and Tang, Yansong},
  booktitle={Proceedings of the AAAI Conference on Artificial Intelligence},
  volume={39},
  number={8},
  pages={8159--8168},
  year={2025}
}

@inproceedings{piczak2015esc,
  title={ESC: Dataset for environmental sound classification},
  author={Piczak, Karol J},
  booktitle={Proceedings of the 23rd ACM international conference on Multimedia},
  pages={1015--1018},
  year={2015}
}

@inproceedings{radford2023robust,
  title={Robust speech recognition via large-scale weak supervision},
  author={Radford, Alec and Kim, Jong Wook and Xu, Tao and Brockman, Greg and McLeavey, Christine and Sutskever, Ilya},
  booktitle={International conference on machine learning},
  pages={28492--28518},
  year={2023},
  organization={PMLR}
}

@article{beltagy2020longformer,
  title={Longformer: The long-document transformer},
  author={Beltagy, Iz and Peters, Matthew E and Cohan, Arman},
  journal={arXiv preprint arXiv:2004.05150},
  year={2020}
}

@article{hsu2021hubert,
  title={Hubert: Self-supervised speech representation learning by masked prediction of hidden units},
  author={Hsu, Wei-Ning and Bolte, Benjamin and Tsai, Yao-Hung Hubert and Lakhotia, Kushal and Salakhutdinov, Ruslan and Mohamed, Abdelrahman},
  journal={IEEE/ACM transactions on audio, speech, and language processing},
  volume={29},
  pages={3451--3460},
  year={2021},
  publisher={IEEE}
}

@inproceedings{milletari2016v,
  title={V-net: Fully convolutional neural networks for volumetric medical image segmentation},
  author={Milletari, Fausto and Navab, Nassir and Ahmadi, Seyed-Ahmad},
  booktitle={2016 fourth international conference on 3D vision (3DV)},
  pages={565--571},
  year={2016},
  organization={Ieee}
}

@inproceedings{xie2015holistically,
  title={Holistically-nested edge detection},
  author={Xie, Saining and Tu, Zhuowen},
  booktitle={Proceedings of the IEEE international conference on computer vision},
  pages={1395--1403},
  year={2015}
}

@inproceedings{devlin2019bert,
  title={Bert: Pre-training of deep bidirectional transformers for language understanding},
  author={Devlin, Jacob and Chang, Ming-Wei and Lee, Kenton and Toutanova, Kristina},
  booktitle={Proceedings of the 2019 conference of the North American chapter of the association for computational linguistics: human language technologies, volume 1 (long and short papers)},
  pages={4171--4186},
  year={2019}
}

@inproceedings{wang2024ref,
  title={Ref-avs: Refer and segment objects in audio-visual scenes},
  author={Wang, Yaoting and Sun, Peiwen and Zhou, Dongzhan and Li, Guangyao and Zhang, Honggang and Hu, Di},
  booktitle={European Conference on Computer Vision},
  pages={196--213},
  year={2024},
  organization={Springer}
}

@article{xiao2025towards,
  title={Towards visual grounding: A survey},
  author={Xiao, Linhui and Yang, Xiaoshan and Lan, Xiangyuan and Wang, Yaowei and Xu, Changsheng},
  journal={IEEE Transactions on Pattern Analysis and Machine Intelligence},
  year={2025},
  publisher={IEEE}
}

@inproceedings{liu2021swin,
  title={Swin transformer: Hierarchical vision transformer using shifted windows},
  author={Liu, Ze and Lin, Yutong and Cao, Yue and Hu, Han and Wei, Yixuan and Zhang, Zheng and Lin, Stephen and Guo, Baining},
  booktitle={Proceedings of the IEEE/CVF international conference on computer vision},
  pages={10012--10022},
  year={2021}
}

@article{lyu2020uavid,
  title={UAVid: A semantic segmentation dataset for UAV imagery},
  author={Lyu, Ye and Vosselman, George and Xia, Gui-Song and Yilmaz, Alper and Yang, Michael Ying},
  journal={ISPRS journal of photogrammetry and remote sensing},
  volume={165},
  pages={108--119},
  year={2020},
  publisher={Elsevier}
}

@inproceedings{radford2021learning,
  title={Learning transferable visual models from natural language supervision},
  author={Radford, Alec and Kim, Jong Wook and Hallacy, Chris and Ramesh, Aditya and Goh, Gabriel and Agarwal, Sandhini and Sastry, Girish and Askell, Amanda and Mishkin, Pamela and Clark, Jack and others},
  booktitle={International conference on machine learning},
  pages={8748--8763},
  year={2021},
  organization={PmLR}
}

@article{yao2025falcon,
  title={Falcon: A remote sensing vision-language foundation model (Technical Report)},
  author={Yao, Kelu and Xu, Nuo and Yang, Rong and Xu, Yingying and Gao, Zhuoyan and Kitrungrotsakul, Titinunt and Ren, Yi and Zhang, Pu and Wang, Jin and Wei, Ning and others},
  journal={arXiv preprint arXiv:2503.11070},
  year={2025}
}

@inproceedings{muhtar2024lhrs,
  title={Lhrs-bot: Empowering remote sensing with vgi-enhanced large multimodal language model},
  author={Muhtar, Dilxat and Li, Zhenshi and Gu, Feng and Zhang, Xueliang and Xiao, Pengfeng},
  booktitle={European Conference on Computer Vision},
  pages={440--457},
  year={2024},
  organization={Springer}
}

@article{zhang2024earthgpt,
  title={EarthGPT: A universal multimodal large language model for multisensor image comprehension in remote sensing domain},
  author={Zhang, Wei and Cai, Miaoxin and Zhang, Tong and Zhuang, Yin and Mao, Xuerui},
  journal={IEEE Transactions on Geoscience and Remote Sensing},
  volume={62},
  pages={1--20},
  year={2024},
  publisher={IEEE}
}

@inproceedings{kuckreja2024geochat,
  title={Geochat: Grounded large vision-language model for remote sensing},
  author={Kuckreja, Kartik and Danish, Muhammad Sohail and Naseer, Muzammal and Das, Abhijit and Khan, Salman and Khan, Fahad Shahbaz},
  booktitle={Proceedings of the IEEE/CVF conference on computer vision and pattern recognition},
  pages={27831--27840},
  year={2024}
}

@article{zhu2022detection,
  title={Detection and tracking meet drones challenge},
  author={Zhu, Pengfei and Wen, Longyin and Du, Dawei and Bian, Xiao and Fan, Heng and Hu, Qinghua and Ling, Haibin},
  journal={IEEE transactions on pattern analysis and machine intelligence},
  volume={44},
  number={11},
  pages={7380--7399},
  year={2021},
  publisher={IEEE}
}

@inproceedings{yao2025uemm,
  title={Uemm-air: Enable uavs to undertake more multi-modal tasks},
  author={Yao, Liang and Liu, Fan and Xu, Shengxiang and Zhang, Chuanyi and Di, Shimin and Ma, Xing and Jiang, Jianyu and Wang, Zequan and Zhou, Jun},
  booktitle={Proceedings of the 33rd ACM International Conference on Multimedia},
  pages={12792--12798},
  year={2025}
}

@inproceedings{yao2026remotereasoner,
  title={Remotereasoner: Towards unifying geospatial reasoning workflow},
  author={Yao, Liang and Liu, Fan and Lu, Hongbo and Zhang, Chuanyi and Min, Rui and Xu, Shengxiang and Di, Shimin and Peng, Pai},
  booktitle={Proceedings of the AAAI Conference on Artificial Intelligence},
  volume={40},
  number={14},
  pages={11883--11891},
  year={2026}
}

@article{liu2025boost,
  title={Boost UAV-based object detection via scale-invariant feature disentanglement and adversarial learning},
  author={Liu, Fan and Yao, Liang and Zhang, Chuanyi and Wu, Ting and Zhang, Xinlei and Jiang, Xiruo and Zhou, Jun},
  journal={IEEE Transactions on Geoscience and Remote Sensing},
  volume={63},
  pages={1--13},
  year={2025},
  publisher={IEEE}
}

@article{liu2026heterogeneous,
  title={Heterogeneous Knowledge Distillation Fostered Pre-training for remote sensing object detection},
  author={Liu, Fan and Zhang, Xinlei and Zhang, Chuanyi and Yao, Liang and Cheng, Fan and Li, Xinyuan and Yu, Xiaohan},
  journal={Pattern Recognition},
  pages={114030},
  year={2026},
  publisher={Elsevier}
}

@article{BO202639,
title = {SARU: A Shadow-Aware and Removal Unified Framework for remote sensing images with new benchmarks},
journal = {ISPRS Journal of Photogrammetry and Remote Sensing},
volume = {239},
pages = {39-53},
year = {2026},
issn = {0924-2716},
doi = {https://doi.org/10.1016/j.isprsjprs.2026.05.022},
url = {https://www.sciencedirect.com/science/article/pii/S0924271626002637},
author = {Zi-Yang Bo and Wei Lu and Hongruixuan Chen and Si-Bao Chen and Bin Luo},
}

@article{LU2026431,
title = {UnravelNet: A backbone for enhanced multi-scale and low-quality feature extraction in remote sensing object detection},
journal = {ISPRS Journal of Photogrammetry and Remote Sensing},
volume = {231},
pages = {431-442},
year = {2026},
issn = {0924-2716},
doi = {https://doi.org/10.1016/j.isprsjprs.2025.11.002},
url = {https://www.sciencedirect.com/science/article/pii/S0924271625004319},
author = {Wei Lu and Hui-Dong Li and Chao Wang and Si-Bao Chen and Chris H.Q. Ding and Jin Tang and Bin Luo}
}

@ARTICLE{RemoteAgent,
  author={Yao, Liang and Xu, Shengxiang and Liu, Fan and Zhang, Chuanyi and Yao, Bishun and Min, Rui and Li, Yongjun and Ouyang, Chaoqian and Di, Shimin and Zhang, Min-Ling},
  journal={IEEE Transactions on Geoscience and Remote Sensing}, 
  title={RemoteAgent: Bridging Vague Human Intents and Earth Observation With RL-Based Agentic MLLMs}, 
  year={2026},
  volume={64},
  number={},
  pages={5636613-5636613},
  doi={10.1109/TGRS.2026.3724339}}

@article{chen2026evaluating,
  title={Evaluating Remote Sensing Image Captions Beyond Metric Biases},
  author={Chen, Ziyun and Liu, Fan and Yao, Liang and Zhang, Chuanyi and Ma, Yuye and Zhou, Wei},
  journal={arXiv preprint arXiv:2604.22855},
  year={2026}
}

@ARTICLE{10345690,
  author={Cho, Yubin and Yu, Hyunwoo and Kang, Suk-Ju},
  journal={IEEE Transactions on Multimedia}, 
  title={Cross-Aware Early Fusion With Stage-Divided Vision and Language Transformer Encoders for Referring Image Segmentation}, 
  year={2024},
  volume={26},
  number={},
  pages={5823-5833},
  doi={10.1109/TMM.2023.3340062}}

@ARTICLE{10227590,
  author={Ding, Haixin and Zhang, Shengchuan and Wu, Qiong and Yu, Songlin and Hu, Jie and Cao, Liujuan and Ji, Rongrong},
  journal={IEEE Transactions on Multimedia}, 
  title={Bilateral Knowledge Interaction Network for Referring Image Segmentation}, 
  year={2024},
  volume={26},
  number={},
  pages={2966-2977},
  doi={10.1109/TMM.2023.3305869}}

@ARTICLE{10413654,
  author={Wang, Wenxuan and He, Xingjian and Zhang, Yisi and Guo, Longteng and Shen, Jiachen and Li, Jiangyun and Liu, Jing},
  journal={IEEE Transactions on Multimedia}, 
  title={CM-MaskSD: Cross-Modality Masked Self-Distillation for Referring Image Segmentation}, 
  year={2024},
  volume={26},
  number={},
  pages={6906-6916},
  doi={10.1109/TMM.2024.3358085}}

@ARTICLE{10510606,
  author={Xu, Sen and Wei, Shikui and Ruan, Tao and Liao, Lixin and Zhao, Yao},
  journal={IEEE Transactions on Multimedia}, 
  title={Each Performs Its Functions: Task Decomposition and Feature Assignment for Audio-Visual Segmentation}, 
  year={2024},
  volume={26},
  number={},
  pages={9489-9498},
  doi={10.1109/TMM.2024.3394682}}

@ARTICLE{10812843,
  author={Shi, Zhaofeng and Wu, Qingbo and Meng, Fanman and Xu, Linfeng and Li, Hongliang},
  journal={IEEE Transactions on Multimedia}, 
  title={Cross-Modal Cognitive Consensus Guided Audio–Visual Segmentation}, 
  year={2025},
  volume={27},
  number={},
  pages={209-223},
  doi={10.1109/TMM.2024.3521746}}

\end{document}